\documentclass[10pt,aps,prd,twocolumn,showpacs,superscriptaddress]{revtex4-2} 

\usepackage{amsmath}
\usepackage{amsfonts}
\usepackage{amssymb}

\newcommand{\beq}{\begin{equation}} \newcommand{\eeq}{\end{equation}}
\newcommand{\bea}{\begin{eqnarray}} \newcommand{\eea}{\end{eqnarray}}

\usepackage[T1]{fontenc}
\usepackage[utf8]{inputenc}
\usepackage[english]{babel}

\usepackage{color}
\usepackage{hyperref}
\hypersetup{colorlinks=true, linkcolor=blue, citecolor=blue, filecolor=blue, urlcolor=blue,pdfproducer={Latex},pdfcreator={pdflatex}}

\newcommand{\UNICV}{\affiliation{Faculdade de Ci\^encias e Tecnologias, Universidade de Cabo Verde, 
7943-010 Praia, Santiago, Cabo Verde
}}
\newcommand{\UFES}{\affiliation{Núcleo Cosmo-ufes \& Departamento de Física, Universidade Federal do Espírito Santo, Avenida Fernando Ferrari 514, 29075-910 Vitória, ES, Brazil
}}
\newcommand{\UFABC}{\affiliation{Centro de Ci\^encias Naturais e Humanas, Universidade Federal
do ABC,  Avenida dos Estados 5001, 09210-580 Santo Andr\'e, São Paulo, Brazil}}

\begin{document}

\title{A Lagrangian formulation for Rastall gravity and a covariant formulation for unimodular gravity}

\author{Juilson A. P. Paiva}\UNICV

\author{Julio C. Fabris}\UFES

\author {Vilson T. Zanchin}\UFABC

\begin{abstract}
We propose a Lagrangian formulation for a non-conservative gravity model in which the divergence of the energy-momentum tensor in curved spacetime does not vanish.
This is accomplished by introducing an arbitrary vector field that couples with the gradient of the Ricci curvature scalar. We first derive the field equations using the Palatini variational approach. Because the connection and the metric tensor are independent in the Palatini framework, the auxiliary vector field dictates whether the manifold geometry is Weyl or Riemannian. By assuming certain physically reasonable conditions on this vector field, the resulting field equations reduce to those of Rastall gravity. Furthermore, slightly different conditions on the vector field furnish unimodular gravity. For comparison, we also employ the standard metric variational approach to obtain the field equations, demonstrating that the same models can be recovered under appropriate conditions. Our key results are the derivation of a covariant Lagrangian formulation for Rastall gravity and a new Lagrangian formulation for unimodular gravity.

\end{abstract}

\maketitle

\section{Introduction}
Since the emergence of cosmological problems in the standard model such as dark energy \cite{Dark-E} and dark matter \cite{Dark-M}, several modified gravity theories have been proposed in order to solve them. Among them, Rastall gravity that was proposed in 1972 \cite{Rast} is the one that most closely matches our solar system tests, besides that, the theory is in agreement with the Einstein's equivalence principle. This is a direct consequence that the vacuum solutions remains the same as in general relativity (GR). The main difference  between general relativity and Rastall gravity consists in the violation  of energy-momentum tensor conservation in curved space-time. Rastall gravity is also in agreement with observational data on the Universe and the Hubble parameter \cite{Hubble}, helium nucleosynthesis \cite{Helium}  and gravitational lensing phenomena \cite{Lens}.  
Moreover, in Ref. \cite{cosmo} it has been shown that a cosmological model based on the Rastall theory reproduces  the success of the $\Lambda$CDM model for the background and at linear perturbative level. It has also been evaluated recently that Rastall-like gravity theories are consistent with the astrophysics of compact objects (see, e.g., \cite{daSilva:2020okh, Vanzella:2022xds,Pattersons:2024fdz}).

Although Rastall gravity presents a compelling alternative to general relativity, it faces notable challenges, including a lack of a full Lagrangian formulation and an inability to describe dark energy in the same manner as standard GR. In this respect, it is worth mentioning the modified Lagrangian proposed in Ref.~\cite{smalley1984}, the 4--index gravity theory formulation of Ref.~\cite{4Rast}, and the tentative Lagrangian formulations for the Rastall gravity in the context of $f(R,T)$ gravity theory as in \cite{5Rast,Shabani:2020wja,epjp}.
In particular, studies such as \cite{4Rast} and \cite{5Rast} have proposed Lagrangian frameworks for the theory by relaxing the stationary condition and restricting the energy-momentum tensor, respectively. However, since a consistent quantized version of the theory requires a rigorous Lagrangian foundation, developing such a formulation remains crucial.

Unimodular gravity, originally proposed by Einstein in 1919 \cite{Einst1919} as a potential avenue to explain the internal structure of matter, breaks away from the standard general relativity framework by fixing the determinant of the metric to a constant value or a specified background volume form, typically chosen as $\sqrt{-g} = 1$. 
As a consequence of this constraint, the theory lacks full general covariance, and the full diffeomorphism group is restricted to transverse (volume-preserving) diffeomorphisms. 
With this restriction, the field equations derived from the unimodular variational principle correspond to the trace-free part of the standard Einstein equations. A crucial feature of this formulation is that the 
cosmological constant $\Lambda$ emerges naturally as an integration constant from the conservation of the energy-momentum tensor via the Bianchi identities. This conceptual shift provides a compelling framework that can partially resolve, or at least significantly alleviate, the cosmological constant problem \cite{Smolin}, as quantum vacuum fluctuations do not automatically couple to gravity in the conventional manner (see also \cite{Bufalo:2015wda}).

While unimodular gravity can be viewed as a specific constrained case of Rastall gravity \cite{Visser:2017gpz}, wherein the divergence of the energy-momentum tensor is proportional to the gradient of the Ricci scalar, their 
structural foundations differ fundamentally. Unlike the general Rastall framework, well-defined and covariant Lagrangian formulations have been successfully developed for unimodular gravity. These are typically achieved by introducing a Lagrange multiplier or utilizing a Henneaux-Teitelboim \cite{teitelboim}
parameterized formalism (see, for instance, \cite{Smolin,Bufalo:2015wda, Saez-Gomez:2016gum, Bengochea:2023dep}). For a comprehensive review of the theoretical status, conceptual challenges, and observational constraints of unimodular gravity, see the status report \cite{Carballo-Rubio:2022ofy}.

In this paper, we establish a novel, standard Lagrangian formulation for Rastall gravity, demonstrating that unimodular gravity naturally emerges as a direct consequence of this fully covariant framework. By utilizing both the Palatini (metric-affine) variation and, independently, the standard metric approach, we provide a robust dual-perspective derivation of both theories. This unified framework not only reconciles the traditionally non-conservative nature of Rastall gravity with a rigorous variational principle but also illuminates a deeper connection between the Rastall and unimodular paradigms.

The text is organized as follows. In next section we propose an action leading to Rastall and unimodular theories. In section \ref{sec:palatini}, the field equations are derived using the Palatini formalism, while in section \ref{sec:metricapproach} the metric approach is used. 
We present our concluding remarks in section \ref{sec:conclusion}. Some technical details are presented in appendices \ref{sec:appendixa}. 

\section{ The postulated action}
\label{se:action}

In the present analysis, we consider the following action
\begin{align}
S =\frac{1}{2k}\int_\Omega\sqrt{-g}\left(R+ a^\mu \nabla_\mu\left(R\right)+2k\mathcal{L}_m\right)d^4x, \label{eq:action} 
\end{align}
where $k$ is the gravitational coupling constant, $k=\frac{8\pi G}{c^2}$, with $G$ being the gravitational constant and $c$ the speed of light, $g$ is the determinant of the metric tensor,  $R=R_{\mu\nu}g^{\mu\nu}$ is the Ricci scalar, with $R_{\mu\nu}$ being the Ricci tensor,  $a^\mu$ is an arbitrary external (background) vector field, and $\mathcal{L}_m$ stands for the Lagrangian density of matter and other non-gravitational dynamical fields. As usual the integration is taken over a volume $\Omega$ of the manifold whose boundary is denoted $\delta\Omega$.

The presence of the term $a^\mu \nabla_\mu R$ in the Lagrangian is justified by the fact that most of dissipative systems are described by odd-order differential equations. In a mechanical system, this can be accomplished by including terms in the Lagrangian with couplings between the fields and their derivatives, so that the resulting equations of motion contain linear first-order or third-order derivatives of the fundamental fields. In the GR theory, the basic assumptions of Riemannian geometry prevent the first derivative of the metric of carrying physical and geometric meaning, since by using local Gaussian coordinates the first-order derivatives of the metric field vanish. Hence, the modification of the Einstein-Hilbert gravitational action capable of generating field equations with relevant odd-order derivatives is by including derivatives of the curvature in the Lagrangian density. The simplest term of this kind is probably one containing the derivative of the Ricci scalar, as we considered in Eq.~\eqref{eq:action}.

Unlike the standard Einstein-Aether theory \cite{eter}, the vector field $a^\mu$ introduced in Eq.~\eqref{eq:action} is non-dynamical, meaning it does not possess its own kinetic term or propagate degrees of freedom. In this respect, the present setup closely resembles the ponderable aether theory \cite{Jacobson:2015mra,Speranza:2015sta}. While this parallel is striking, a detailed, quantitative comparison between these two specific frameworks lies beyond the scope of the present work and is left for future investigation.

Notice that the action \eqref{eq:action} may be written in the form 
\begin{align}
S = \frac{1}{2k}\!\int_\Omega\!\!\sqrt{-g}\Big(R\left[1-\nabla_\mu a^\mu\right]+\nabla_\mu\left(a^\mu R\right) +2k\mathcal{L}_m\Big)d^4x. \label{eq:action-b}
\end{align}
This form is suited for the Palatini method, whereas Eq.~\eqref{eq:action} is better suited for the standard metric approach. See below for details.

\section{Field equations from the Palatini variational method }
\label{sec:palatini}

\subsection{General field equations}

 The Palatini variational method for the gravitational interaction assumes that the metric tensor $ g_{\mu\nu} $ and connection $ \Gamma_{\mu \nu}^\rho $ are fundamental dynamical fields of the theory. In such a case, the metric and the connection are independent quantities, i.e., in principle there is no relationship between them, and the complete set of field equations is obtained by varying the action with respect to both the metric and the connection.

 Let us first consider the variation of the action \eqref{eq:action} with respect to the metric inverse $g^{\mu\nu}$, where it vanishes on the boundary surface. Thus, we have
 \begin{align}
\delta S = &\frac{1}{2k}\!\int_\Omega\!\sqrt{-g}\left[\bar G_{\mu\nu}-\frac{1}{2}g_{\mu\nu}\nabla_{\!\rho}\!\left(a^\rho R\right) -kT_{\mu\nu}\right]\delta g^{\mu\nu}d^4x, \nonumber \\
&+\frac{1}{2k}\int_\Omega\!\sqrt{-g}\,\nabla_\rho\big(a^\rho R_{\mu\nu}\, \delta g^{\mu\nu}\big)d^4x, \label{eq:var1} 
 \end{align}
where we defined 
\begin{equation}
  \tilde G_{\mu\nu} = \left(1 -\nabla_\rho a^\rho\right)G_{\mu\nu},
\end{equation}
with $G_{\mu\nu}$ being the Einstein tensor, and $T_{\mu\nu}$ is
the energy-momentum tensor, given by the usual expression 
\begin{equation}
    T_{\mu\nu}=-\frac{2}{\sqrt{-g}}\frac{\delta\left(\mathcal{L}_m\sqrt{-g}\right)}{\delta g^{\mu\nu}}.
\end{equation}

The last term in \eqref{eq:var1} does not contribute because $\delta g^{\mu\nu}$ vanishes at the boundary. Specifically, applying the divergence theorem transforms the volume integral into a surface integral over the boundary,
\begin{equation}
\begin{split}
& \int_\Omega \sqrt{-g\,}\,\nabla_\rho\left(a^\rho R_{\mu\nu}, \delta g^{\mu\nu}\right)d^4x \\
& \qquad\qquad = \int_{\partial\Omega} \sqrt{-g\,}\left(a^\rho R_{\mu\nu}, \delta g^{\mu\nu}\right)d\Sigma_\rho=0.
\end{split}
\end{equation}
Hence, by demanding the action to be stationary, it follows
\begin{equation}
 \left(1-\nabla_\rho a^\rho\right)G_{\mu\nu}-\frac{1}{2}g_{\mu\nu}\nabla_\rho\left(a^\rho R\right) =kT_{\mu\nu}. \label{eq:fieldpalatini1}
\end{equation}
These are the field equations obtained by varying the action \eqref{eq:action} with respect to the metric. To complete the system of equations according to the Palatini formalism, we must also vary the action with respect to the connection. 

Varying the action \eqref{eq:action} with respect to the connection \(\Gamma _{\mu \nu }^{\rho }\)
yields
\begin{equation}
\begin{split}
\delta S&=\int_\Omega\Big[\tilde{g}^{\mu\nu}\delta R_{\mu\nu}+\nabla_\rho\left(a^\rho g^{\mu\nu}\delta R_{\mu\nu} \right)\Big]\sqrt{-g}\,d^4x\\
&=\int_\Omega \tilde{g}^{\mu\nu}\delta R_{\mu\nu}\sqrt{-g}\,d^4x +\int_{\partial\Omega}\left(a^\rho g^{\mu\nu}\delta R_{\mu\nu} \right)d\Sigma_\rho, \\ 
&=\int_\Omega\Big[\nabla_\nu\left(\delta\Gamma^\sigma_{\mu\sigma}\right)-\nabla_\sigma\left(\delta\Gamma^\sigma_{\mu\nu}\right)\Big]\tilde{g}^{\mu\nu}\sqrt{-g}\,d^4x \label{eq:var-conn} \\  
&\;\; +\int_{\partial\Omega} a^\rho g^{\mu\nu}\Big[\partial_\nu\left(\delta\Gamma^\sigma_{\mu\sigma}\right)-\partial_\sigma\left(\delta\Gamma^\sigma_{\mu\nu} \right)\Big] d\Sigma_\rho, \\
\end{split}
\end{equation}
where to simplify notation we introduced the modified metric
\begin{equation}
    \tilde{g}^{\mu\nu}=\big[1-\left(\nabla_\rho a^\rho\right)\big]g^{\mu\nu}.
\end{equation}

The surface integral in \eqref{eq:var-conn} vanishes just by assuming that the derivatives of the connection variations $\partial_\rho\left[\delta\Gamma^\sigma_{\mu\nu}\right]$ vanish at the boundary $\delta\Omega$. The usual way to avoid this extra-assumption is to add counter-terms in the original Lagrangian. 

By performing an integration by parts and relabeling some dummy indices, Eq.~\eqref{eq:var-conn} can be recast as
\begin{equation}
\begin{split}
\delta S=
&\int_\Omega\nabla_\nu\left[\left(\tilde{g}^{\mu\nu}\delta\Gamma^\sigma_{\mu\sigma}-\tilde{g}^{\mu\sigma}\delta\Gamma^\nu_{\mu\sigma}\right)\right]\sqrt{-g}\,d^4x\\
&-\int_\Omega\left[\nabla_\rho\tilde{g}^{\mu\nu}\delta\Gamma^\rho_{\mu\nu}-\nabla_\nu\tilde{g}^{\mu\nu}\delta\Gamma^\rho_{\mu\rho}\right]\sqrt{-g}\,d^4x.
\end{split}
\end{equation}
 Notice that the metric compatibility with the connection was not used here, as we have assumed that the metric and the connection are independent fields. Another integration by parts in the first expression transforms the integral on the volume $ \Omega $ into a surface integral over $ \delta \Omega $. Since the connection variation is null in $ \delta \Omega $, we can ignore it. Therefore, from the freedom that we have to relabel dummy indices, we get
\begin{equation}
\delta S=\int_\Omega\left[\delta_\rho^\nu\nabla_\sigma\tilde{g}^{\mu\sigma}-\nabla_\rho\tilde{g}^{\mu\nu}\right]\delta\Gamma^\rho_{\mu\nu}\sqrt{-g}d^4x.\label{eq7}
\end{equation}
Taking into account that any tensor can be decomposed into a  symmetric and  anti-symmetric part and that the connection is torsion free, i.e., $ \Gamma^\rho_{\mu \nu} = \Gamma^\rho_{\nu \mu} $, the variation \eqref{eq7} reduces to
\begin{equation}
\begin{split}
\delta S=&\int_\Omega\left[\frac{1}{2}\delta_\rho^\nu\nabla_\sigma\tilde{g}^{\mu\sigma}+\frac{1}{2}\delta_\rho^\mu\nabla_\sigma\tilde{g}^{\nu\sigma}\right]\delta\Gamma^\rho_{\mu\nu}\sqrt{-g}d^4x\\
-&\int_\Omega\nabla_\rho\tilde{g}^{\mu\nu}\delta\Gamma^\rho_{\mu\nu}\sqrt{-g}d^4x,
\end{split}
\end{equation}
since the contraction between the anti-symmetric part and $ \delta \Gamma^\rho_{\mu \nu} $ is null. Finally, by imposing the stationary  condition, we obtain the equation
\begin{equation}
\frac{1}{2}\delta_\rho^\nu\nabla_\sigma \tilde{g}^{\mu\sigma}+\frac{1}{2}\delta_\rho^\mu\nabla_\sigma\tilde{g}^{\nu\sigma}-\nabla_\rho\tilde{g}^{\mu\nu}=0.
\end{equation}
Based on the definition of the Kronecker delta $\delta^\mu_\rho$ tensor, it follows that $\nabla_\rho\tilde{g}^{\mu\nu}=0$. Consequently, we find the relation
\begin{equation}
\nabla_\rho g_{\mu\nu}= \frac{\partial_\rho \left(\nabla_\sigma a^\sigma\right)}{1-\left(\nabla_\sigma a^\sigma\right)}\,g_{\mu\nu}.\label{eq10}
\end{equation}
After some algebraic manipulation, we obtain the following connection:
\begin{equation}
\begin{split}
\Gamma_{\mu\nu}^\rho=&\frac{1}{2}g^{\rho\sigma}\left(\partial_\mu g_{\sigma\nu}+\partial_\nu g_{\mu\sigma}-\partial_\sigma g_{\mu\nu}\right)\\
&-\frac{1}{2}\left(\omega_\mu\delta_\nu^\rho+\omega_\nu\delta_\mu^\rho-g_{\mu\nu}\omega^\rho\right),\label{eq:connWeyl}
\end{split}
\end{equation}
where 
\begin{equation}
    \omega_\mu =\frac{\partial_\mu\left(\nabla_\rho a^\rho\right)}{1- \left(\nabla_\rho a^\rho\right)}= -\partial_\mu\big[\ln\left(1-\nabla_\rho a^\rho\right)\big]. \label{eq:weyl1form}
\end{equation}

A geometry defined by a metric satisfying condition \eqref{eq10} corresponds to an integrable Weyl geometry \cite{Weyl1,Weyl2,Einstein}, in which the Weyl 1-form $\omega_\mu$ is an exact differential. As is well known, in the integrable case this effect is purely a gauge artifact, in the sense that the Weyl field can be eliminated by a suitable conformal transformation that maps the geometry to a Riemannian one with a Levi-Civita connection. However, to simplify the discussion and focus on the primary objective of the present analysis, we impose the condition
\begin{equation}
\partial_\mu \left(\nabla_\rho a^\rho\right) = 0,
\label{eq:constphi}
\end{equation}
in order to make the Weyl 1-form \eqref{eq:weyl1form} vanish.
Under this constraint, the Weyl connection \eqref{eq:connWeyl} reduces to the Levi-Civita connection, 
\begin{equation}
\Gamma_{\mu\nu}^\rho=\frac{1}{2}g^{\rho\sigma}\left(\partial_\mu g_{\sigma\nu}+\partial_\nu g_{\mu\sigma}-\partial_\sigma g_{\mu\nu}\right).
\end{equation}

As demonstrated, the vector field  $a^\mu$ is the fundamental element that determines whether the underlying geometry is integrable-Weyl or Riemannian.

Finally, by imposing condition \eqref{eq:constphi} on $a^\mu$, the field equations derived from the action \eqref{eq:action} via the Palatini approach, written in Eq.~\eqref{eq:fieldpalatini1}, may be recast into the form 
\begin{align}
&G_{\mu\nu}-\left(\nabla_\rho a^\rho\right) R_{\mu\nu}-\frac{1}{2}g_{\mu\nu}a^\rho\nabla_\rho R =kT_{\mu\nu}, \label{eq:fieldPalatini1}
\end{align}
with the subsidiary condition $\nabla_\mu a^\mu =\, {\rm constant}$, that follows from Eq.~\eqref{eq:connWeyl}. 

In general, in curved spacetimes satisfying the field equations \eqref{eq:constphi} and  \eqref{eq:fieldPalatini1} the energy-momentum tensor is not conserved, that is $\nabla_\nu T^\nu_{\;\mu}\neq0$. Hence, this theory belongs to the class of Rastall gravity theories. The field equations \eqref{eq:fieldPalatini1} produce third order non-linear differential equations, this would be expected since the action \eqref{eq:action} has third order derivatives in the metric tensor.

It is important to note that the Einstein field equations are recovered by taking $ a^\mu = 0 $. More generally, we can show that Einstein gravity also results in the case $ a^\mu $ belongs to the Killing vector subspace. In fact, if $ a^\mu $ is a Killing vector, we have
\begin{equation}
\nabla_\mu a_\nu+\nabla_\nu a_\mu=0,
\end{equation} 
which after contraction gives us $ \nabla_\mu a^\mu = 0 $. This means that the Einstein gravity also results in the case $a^\mu$ is a constant vector field. It can also be shown that if $ a^\mu $ is a Killing vector the identity $a^\mu\nabla_\mu R=0$ olds, and equation \eqref{eq:fieldPalatini1} reduces to the Einstein field equations. 

Since $\nabla_\mu a^\mu $ is a constant, and  defining 
\begin{equation}\label{eq:phi}
    \phi\equiv \nabla_\mu a^\mu, 
\end{equation}
Eq.~\eqref{eq:fieldPalatini1} can be written in the alternative form
\begin{equation}
    G_{\mu\nu}-\dfrac{1}{2} g_{\mu\nu}\left(\tilde{\phi}R+\tilde{a}^\rho\nabla_\rho R\right)=\tilde{k}T_{\mu\nu},\label{eq16}
\end{equation}
where we redefined the parameters as follows,
\begin{equation}
    \tilde{\phi}=\frac{\phi}{1-\phi}, \quad \tilde{a}^\mu=\frac{a^\mu}{1-\phi}, \quad \tilde{k}=\frac{k}{1-\phi}, 
\end{equation}
with $\phi\neq 1$.
The particular form of the field equations given by Eq.~\eqref{eq16} is well-suited for our subsequent analysis.

\subsection{Rastall gravity}

Rastall gravity \cite{Rast} is based on the field equations 
$R_{\mu\nu} + \left(k\lambda -1/{2}\right)g_{\mu\nu} R = k T_{\mu\nu}$, with $k\lambda \neq 1/4$. By comparing these equations with Eq.~\eqref{eq:fieldPalatini1} we see that in order to obtain Rastall gravity, we must fix the condition
\begin{equation} \label{eq:rastallcond1}    
a^\mu\nabla_\mu R=0. 
\end{equation}
In principle, this condition can always be satisfied, since the arbitrary vector field $a^\mu $ bears four degrees of freedom, and even after imposing the condition on the field $a^\mu$ resulting from Eq.~\eqref{eq:constphi}, the degrees of freedom are reduced to just three. A practical example is when space-time present some kind of symmetry. This may be illustrated by taking a spherically symmetric space-time considering spherical coordinates. In such a case, it follows $\partial_\theta R=\partial_\varphi R=0$ and, therefore, we can choose $a^\mu=\left(0,0,a^\theta,a^\varphi\right)$ to satisfy the imposed condition.

In conclusion, after substituting the constraint \eqref{eq:rastallcond1} into Eq.~\eqref{eq:fieldPalatini1}, the standard Rastall field equation 
\begin{equation}\label{eq:rastall}
    G_{\mu\nu}-\dfrac{\tilde{\phi}}{2} g_{\mu\nu}R=\tilde{k}T_{\mu\nu}
\end{equation}
is obtained. In the last equation, the constant $\tilde{\phi}$ and the standard Rastall parameter $\lambda$ are related by $\tilde{\phi}=-2\tilde{k}\lambda$.

Note that, if we impose the conservation of the energy-momentum tensor $T_{\mu\nu}$, the cosmological constant emerges as an integration constant, such that $R\big(1 +2\tilde\phi\big) = -2\tilde{k}T = \text{constant}$. In fact, taking the divergence of the field equations \eqref{eq:rastall}, we obtain
\begin{equation}
    \tilde\phi \partial_\mu R = -2\tilde k\nabla_\nu T^\nu_\mu.
\end{equation}
For a non-vanishing $\tilde\phi$, the conservation condition $\nabla_\nu T^\nu_\mu = 0$ implies $\partial_\mu R = 0$, which requires the Ricci $R$ to be an arbitrary constant. Hence, the standard Einstein field equations with a cosmological constant are recovered from the present theory by imposing the conservation of the energy-momentum tensor. 
However, because of the relation $R\big(1 +2\tilde\phi\big) = -2\tilde{k}T$, this correspondence holds specifically for energy-momentum tensors with a constant or vanishing trace. Beyond the cosmological constant case, electrovacuum spacetimes present another interesting scenario where this model yields the same equations as general relativity. Furthermore, taking the limit $\tilde\phi \to 0$ in the divergence condition successfully recovers the full, standard Einstein field equations.

\subsection{Unimodular gravity}

On the other hand, after taking the trace of Eq.~\eqref{eq16}, we may write the trace-free equation
\begin{equation}
    G_{\mu\nu}+\frac{1}{4}g_{\mu\nu}R=\tilde{k}\left(T_{\mu\nu}-\frac{1}{4}g_{\mu\nu}T\right),\label{eq18}
    \end{equation}
with the following auxiliary conditions,
\begin{eqnarray}
        &R\left(1+ 2\tilde \phi\right) +\tilde{k}T+2\tilde{a}^\mu \nabla_\mu R=0,\label{eq:tracepalatini}\\
        & \nabla_\mu\phi=0.\label{eq19b}
\end{eqnarray}
 Notice that the constraint \eqref{eq19b} is the same as the one imposed above to get the Levi-Civita connection within the Palatini approach, cf. Eq.~\eqref{eq:constphi}. Moreover, the condition \eqref{eq:rastallcond1} is not necessary here. 

The field equations \eqref{eq18} are formally identical to those found in unimodular gravity. Originally formulated by Einstein \cite{Einst1919}, these field equations have been independently rederived  multiple times across different contexts \cite{item1,teitelboim,item3,item4}.  As mentioned above, the field equations of unimodular gravity may be derived from the standard Einstein-Hilbert action by explicitly imposing the unimodular constraint $\sqrt{-g}=f$, where $f$ is a fixed scalar density (see, e.g., Refs.~\cite{Saez-Gomez:2016gum,Bengochea:2023dep,Carballo-Rubio:2022ofy}).
Following the original proposal, many authors choose $\sqrt{-g}=1$.
Because this constraint singles out a specific class of coordinate systems, standard unimodular gravity inherently lacks explicit general covariance.
In contrast, our approach circumvents this limitation entirely, it requires no a priori coordinate restrictions and remains fully covariant. Indeed, the standard form of the unimodular gravity equations \eqref{eq18} emerges naturally from the purely geometric constraint presented in Eq.~\eqref{eq:tracepalatini}.
A crucial feature of this relation is the presence of the vector field $a^\mu$. Because $a^\mu$ acts as an arbitrary gauge-like or auxiliary vector, its presence ensures that the relation does not enforce an additional constraint on the Ricci scalar $R$ itself, thereby preserving the full degrees of freedom of the theory.

As with the Rastall-type gravity presented in the preceding section, Einstein equations with a cosmological constant emerge naturally when energy-momentum conservation is imposed. Specifically, by taking the divergence of the field equations \eqref{eq18} and requiring the energy-momentum tensor $T_{\mu\nu}$ to be conserved, the cosmological constant appears as a constant of integration: $R+\tilde{k}T=-2\nabla_\mu\left(\tilde{a}^\mu R\right)={\rm constant}=\Lambda$. 

It is worth also mentioning that, by imposing the Rastall constraint \eqref{eq:rastallcond1}, the derivative term in Eq.~\eqref{eq:tracepalatini} vanishes, yielding, $R(1+2\tilde\phi)+\tilde{k}T=0$.
This is precisely the trace of the Rastall field equations \eqref{eq:rastall}, confirming the internal consistency of our generalized formulation.

\section{Field equations from the metric (Einstein-Hilbert) variational approach}
\label{sec:metricapproach}

\subsection{The general field equations}

In the standard formulation of the Einstein-Hilbert action, the metric tensor $g_{\mu\nu}$ is treated as the sole independent field, and the connection is assumed to be metric-compatible. By varying the action in Eq. \eqref{eq:action} with respect to the metric we arrive at the following field equations,
\begin{equation}
    G_{\mu\nu}-\phi R_{\mu\nu}-\frac{1}{2}g_{\mu\nu}a^\rho\nabla_\rho R+L_{\mu\nu}=kT_{\mu\nu},\label{eq:fieldclassical}
\end{equation}
where $\phi= \left(\nabla_\rho a^\rho\right)$ and 
\begin{equation}
L_{\mu\nu}=\nabla_\mu\nabla_\nu\phi-g_{\mu\nu}\square\phi, \label{eq:Lmunu}
\end{equation}
where the covariant d'Alembert operator, $\square=g^{\mu\nu}\nabla_\mu\nabla_\nu$. For more details on the derivation of \eqref{eq:fieldclassical}, see Appendix \ref{sec:appendixa}.

In flat space-time, Eq.~\eqref{eq:fieldclassical} becomes
\begin{equation}
    kT_{\mu\nu}=\nabla_\mu\nabla_\nu\phi-\eta_{\mu\nu}\square\phi,
\end{equation}
and then, since $\phi$ is a smooth function on the space-time, we get $ 
    \partial_\mu T^\mu_{\;\nu}=0$.
Thus, in flat spacetime the energy-momentum tensor is conserved. 

Here, unlike the theory obtained from the Palatini approach, there is no relationship between the scalar field $\phi$ and the metric tensor $g_{\mu\nu}$, but both theories coincide when $\phi$ is constant, as expected since the connection is  metric-compatible.

\subsection{Rastall gravity}
 
A comparison between the field equations \eqref{eq:fieldclassical} and the original Rastall equations reveals that the terms containing $a^\mu\nabla_\mu R$ and $L_{\mu\nu}$ must be eliminated. In such a case, Eq.~\eqref{eq:fieldclassical} assumes the form
\begin{equation}
    G_{\mu\nu}-\phi R_{\mu\nu} = k T_{\mu\nu}. \label{eq:classicalRastall}
\end{equation}
Then, the standard Rastall equations follow provided that $\phi$ is a constant and related to the standard Rastall parameter $\lambda$ by $\phi = -k\lambda$.

The contribution $-\dfrac{1}{2}g_{\mu\nu}a^\rho\nabla_\rho R$ vanishes if we consider the same condition used to obtain Rastall gravity through the Palatini approach, cf.~Eq.~\eqref{eq:rastallcond1}. In turn, the condition $L_{\mu\nu} = 0$ yields $\nabla_\mu\nabla_\nu\phi - g_{\mu\nu}\square\phi = 0$. The simplest way to satisfy this condition is, in fact, by imposing the same constraint on $\phi$ as in Eq.~\eqref{eq:constphi}, i.e., 
$\phi =$ constant. Let us mention that the most general field $\phi$ that satisfies the condition $L_{\mu\nu}=0$ obeys the condition $\nabla_\mu\nabla_\nu \phi =0$, which means that the vector $\nabla_\mu \phi$ must be a covariantly constant vector field. However, even though this more general, non-globally constant $\phi$ does imply $L_{\mu\nu}=0$, it does not lead to the correct Rastall field equations.

\subsection{Unimodular gravity}

The field equations \eqref{eq:fieldclassical} may be rewritten in the traceless version  
\begin{equation}
    G_{\mu\nu}+\frac{1}{4}g_{\mu\nu} R+\frac{1}{1-\phi}\left(L_{\mu\nu}-\frac{1}{4}g_{\mu\nu}L\right)=\tilde{k}T^{tf}_{\mu\nu}, \label{eq:feildclassiclTF}
\end{equation}
where the trace free energy-momentum tensor is  $T^{tf}_{\mu\nu}=T_{\mu\nu}-\frac{1}{4}g_{\mu\nu}T$, with the additional trace equation
\begin{equation}
    \nabla_\mu\left(a^\mu R\right)=-\frac{1}{4}\left(3\square\phi+kT+\left(1-\phi\right)R\right). \label{eq:addit}
\end{equation}

A comparison between the field equations \eqref{eq:feildclassiclTF} and the original equations for unimodular gravity shows that the present theory reduces to unimodular gravity only if we impose conditions that eliminate the terms containing $L_{\mu\nu}$ and its trace. Furthermore, $\tilde k = k/(1-\phi)$ mus be a constant. As discussed above, it is necessary to set $\phi$ to a constant.
With this condition, Eq.~\eqref{eq:feildclassiclTF} reduces to the standard form of unimodular gravity,
\begin{equation}
    G_{\mu\nu}+\frac{1}{4}g_{\mu\nu}R=\tilde{k}\left(T_{\mu\nu}-\frac{1}{4}g_{\mu\nu}T\right), \label{eq18b}
    \end{equation}
as expected. Additionally, the present theory requires that the  auxiliary conditions
\begin{equation}
    \begin{split}
        &R+\tilde{k}T+2\nabla_\mu\left(\tilde{a}^\mu R\right)=0,\\
        &\nabla_\mu\left(\nabla_\nu a^\nu\right)\equiv \nabla_\mu\phi=0,\label{eq19}
    \end{split}
\end{equation}
must be fulfilled. 

The first constraint in \eqref{eq19} originates from Eq.~\eqref{eq:addit} and is identical to the necessary trace constraint \eqref{eq:tracepalatini} required to obtain the unimodular equations in the Palatini approach. 
In turn, the second constraint provides the sufficient condition for $L_{\mu\nu}$ to vanish, guaranteeing that $\tilde k$ is indeed a constant parameter. 
It is also worth mentioning that this second constraint in \eqref{eq19} matches the condition for the geometry obtained via the Palatini approach to be Riemannian, cf. Eq.~\eqref{eq:constphi}.

The field equations \eqref{eq18b} are identical to \eqref{eq18}. Unlike unimodular gravity obtained from the Einstein-Hilbert action with the subsidiary constraint $\sqrt{-g} = f$, equations \eqref{eq18b} are fully covariant, which mirrors the results derived from the Palatini approach.

The present theory also shares other properties depicted by the previous version of unimodular gravity obtained above by means of the Palatini approach. 
In fact, after taking the divergent covariant in the field equations \eqref{eq18b} and \eqref{eq19}, and imposing the conservation of the energy-momentum tensor $T_{\mu\nu}$, the cosmological constant emerges as an integration constant, that is $R+\tilde{k}T=-2\nabla_\mu\left(\tilde{a}^\mu R\right)=\Lambda$, where $\Lambda$ is an arbitrary constant. Therefore, the standard Einstein field equation with the cosmological constant arises from the conservation  of energy-momentum tensor. Furthermore, in flat space-time the constant $\Lambda$ vanishes, so the gravitational field  does not couple with the vacuum energy in this theory.

\section{Conclusion}
\label{sec:conclusion}

By introducing an arbitrary auxiliary four-vector coupled directly to the gradient of the Ricci scalar within the gravitational Lagrangian, we establish a novel modified gravity framework specifically tailored for non-conservative systems. Utilizing a variational approach, we rigorously derive the corresponding gravitational field equations in explicit covariant form via both the metric and Palatini formalisms. \\
Rastall gravity: We demonstrate that imposing a particular set of constraints on the auxiliary four-vector naturally reproduces the field equations of Rastall gravity. Because these equations are derived directly from an action principle, this work successfully provides a long-sought Lagrangian formulation for Rastall  theory, firmly establishing it as a first-principles framework rather than a purely phenomenological one.\\
Unimodular gravity: Alternatively, by implementing a different, distinct set of constraints, the framework yields the fully covariant field equations of unimodular gravity.

Ultimately, this Lagrangian proposal provides a robust, mathematically consistent bridge connecting non-conservative gravitational dynamics, Rastall gravity, and unimodular gravity within a single variational architecture.
 Within this theory, the local conservation of the energy-momentum tensor dictates that the cosmological constant emerges dynamically as an integration constant, offering a compelling geometric origin for the dark energy sector.

A further investigation we are currently considering is the Hamiltonian formulation for the present Rastall and unimodular gravities. Such a formulation is not straightforward due to the different constraints that have to be taken into account.

\section{Acknowledgements}
 J.F. thanks Conselho Nacional de Desenvolvimento Científico e Tecnológico (CNPq), Brazil, and Fundação de Amparo à Pesquisa e Inovação do Estado do Espírito Santo (FAPES), Brazil, for partial support.  
 V.T.Z. is partially supported by Conselho Nacional de Desenvolvimento Científico e Tecnológico (CNPq), Brazil, Grant No.~311726/2022-4.
 
\appendix
\section{Derivation of the field equations via metric variation}
\label{sec:appendixa}

We initially consider only the gravitational sector of the action \eqref{eq:action}
 \begin{eqnarray}
S_G & = &\dfrac{1}{2\kappa}\int d^4x\sqrt{-g}\bigr(R + a^\rho\nabla_\rho R \bigl). \label{action1} 
\end{eqnarray}
We denote the variation of the metric and its inverse by,
\begin{eqnarray}
\delta g_{\mu\nu} = - h_{\mu\nu}, \quad \delta g^{\mu\nu} = h^{\mu\nu},
\end{eqnarray}
respectively.
Varying the action \eqref{action1} with respect to the metric yields: 
\begin{equation}
\begin{split}
 \delta S_G =& - \dfrac{1}{2\kappa}\int d^4x \sqrt{-g} \,\dfrac{g_{\mu\nu}}{2}\Bigr(R + a^\rho\nabla_\rho R\Bigl)h^{\mu\nu}\\
&
+ \dfrac{1}{2\kappa}\int d^4 x\sqrt{-g}\Bigr[h^{\mu\nu}R_{\mu\nu} + g^{\mu\nu}\delta R_{\mu\nu} \\ 
&+ a^\rho\nabla_\rho\Bigr(h^{\mu\nu}R_{\mu\nu} + g^{\mu\nu}\delta R_{\mu\nu}\Bigl)\Bigl].
\end{split} 
\end{equation}
After an integration by parts of the terms containing $a^\rho$, and defining
\begin{eqnarray}
 \psi = 1 - \nabla_\rho a^\rho \equiv 1- \phi, \label{eq:psi}
\end{eqnarray}
the final expression for $\delta S_G$ is given by:
\begin{eqnarray}
& \displaystyle{ \delta S_G = \dfrac{1}{2\kappa} \int d^4x\sqrt{-g} h^{\mu\nu}\Bigr[\psi R_{\mu\nu} - \frac{1}{2}g_{\mu\nu}\big(R  + a^\rho\nabla_\rho R\big)\Bigl] }\nonumber\\
& \displaystyle{+ \dfrac{1}{2\kappa} \int d^4x \sqrt{-g}\,\psi\, g^{\mu\nu}\,\delta R_{\mu\nu}}\nonumber \\ 
&\displaystyle{ + \dfrac{1}{2\kappa}\int d^4x \sqrt{-g}\,\nabla_{\rho}\Bigr[a^\rho\Bigr(h^{\mu\nu}R_{\mu\nu} + g^{\mu\nu}\delta R_{\mu\nu}\Bigl)\Bigl]}.  \label{eq:var1a}
\end{eqnarray}

The variation of the Ricci tensor is given by
\begin{eqnarray} 
\label{ricci}
g^{\mu\nu}\delta R_{\mu\nu} =  g^{\rho\sigma}g^{\mu\nu}\Big(\nabla_\sigma\nabla_\rho  h_{\mu\nu} - \nabla_\sigma\nabla_\nu h_{\mu\rho} \Big).
\end{eqnarray}
Using (\ref{ricci}), and considering that integration by parts are going to be performed,  we may write the term $\psi\, g^{\mu\nu}\,\delta R_{\mu\nu}$ in the form
\begin{equation} \label{var2a}
\begin{split}
& \psi\, g^{\mu\nu}\,\delta R_{\mu\nu} =  -h^{\mu\nu}\Big(\nabla_\mu\nabla_\nu \psi- g_{\mu\nu}\Box\psi\Big)\\ 
 & +  \nabla_\rho \biggr[\psi \Big(\nabla_\sigma h^{\sigma\rho} - \nabla^\rho h^{\sigma}_\sigma\Big) + \Big( h^{\sigma\rho} - g^{\rho\sigma}h^{\mu}_{\mu}\Big)\nabla_\sigma \psi\bigg].
\end{split}
\end{equation}
Then, substituting (\ref{var2a}) into \eqref{eq:var1a} we obtain, 
\begin{equation} \label{var2}
\begin{split}
&\delta S_G = \frac{1}{2\kappa} \int d^4x\sqrt{-g} h^{\mu\nu}\, \left(\psi R_{\mu\nu} - \frac{1}{2}g_{\mu\nu}R \right. \\  \  &\left. - \frac{1}{2}g_{\mu\nu}a^\rho\nabla_\rho R - \nabla_\mu\nabla_\nu\psi + g_{\mu\nu}\Box\psi\right) \\ 
&- \frac{1}{2\kappa} \int d^4x \sqrt{-g}\nabla_\rho \left[\psi \Big(\nabla_\sigma h^{\sigma\rho} - \nabla^\rho h^{\sigma}_\sigma\Big) \right] \\ & +  \frac{1}{2\kappa} \int d^4x \sqrt{-g}\nabla_\sigma\left[\Big( h^{\sigma\rho} - g^{\rho\sigma}h^{\mu}_{\mu}\Big)\nabla_\rho\psi\right] \\
&+ \frac{1}{2\kappa}\int d^4x \sqrt{-g}\nabla_{\rho}\left[ a^\rho\Big(h^{\mu\nu}R_{\mu\nu} + g^{\mu\nu}\delta R_{\mu\nu}\Big)\right].  
\end{split} 
\end{equation}
Assuming the surface terms vanish at the boundaries, the last three terms in Eq. (8) do not contribute to the equations of motion. Additionally, including the matter component in the Lagrangian yields the field equations
\begin{equation} \begin{split}
& \psi\, R_{\mu\nu} - \frac{1}{2}g_{\mu\nu} R- \frac{1}{2}g_{\mu\nu} a^\rho \nabla_\rho R \\ & \hspace*{2.6cm} - \nabla_\mu\nabla_\nu \psi + g_{\mu\nu}\Box\psi = \kappa\, T_{\mu\nu} 
\end{split}
\end{equation}
which are identical to Eq.~\eqref{eq:fieldclassical}. Furthermore, taking $\psi =$ constant, defining
\begin{eqnarray}
\lambda = \frac{1}{\psi}, \quad \frac{\kappa}{\psi} = 8\pi G,
\end{eqnarray}
and imposing
\begin{eqnarray}
\label{cond1}
a^\rho\nabla_\rho R = 0, \label{eq:diffcond}
\end{eqnarray}
the Rastall equations are found: 
\begin{eqnarray}
R_{\mu\nu} - \frac{\lambda}{2}g_{\mu\nu} R = 8\pi GT_{\mu\nu}.
\end{eqnarray}

After the assumptions $\psi = $ constant the variation of the total Lagrangian \eqref{eq:action} can be cast into the form
\begin{equation} \label{eq:var3}
\begin{split}
\delta S & = \frac{1}{2\kappa} \int d^4x\sqrt{-g} h^{\mu\nu}\,\left( G_{\mu\nu} - \frac{\phi}{2}g_{\mu\nu}R \right. \\ & \left. - \frac{1}{2}g_{\mu\nu}a^\rho\nabla_\rho R - \kappa T_{\mu\nu}\right) +\, \textrm{ST},  
\end{split} 
\end{equation}
where ST stands for surface terms. 

Considering an infinitesimal diffeomorphism generated by the vector field $\xi^\mu$, under the coordinate transformation $x^\mu \to x^\mu +\xi^\mu$, the metric tensor transforms as $ h^{\mu\nu}\equiv - {\cal L}_\xi g^{\mu\nu} = \nabla^\mu\xi^\nu + \nabla^\nu\xi^\mu$. Substituting $h^{\mu\nu}$ into Eq. \eqref{eq:var3}, we obtain
\begin{equation} \label{var4}
\begin{split}
\delta S_\xi & = \frac{1}{\kappa} \int d^4x\sqrt{-g}\big(\nabla^\nu \xi^\mu\big) \left(G_{\mu\nu} - \frac{1}{2} g_{\mu\nu}\, \phi\, R \right. \\ &  \left.- \frac{1}{2}g_{\mu\nu}\big(a^\rho\nabla_\rho R\big) - \kappa T_{\mu\nu}\right) +\, \textrm{ST}, 
\end{split} 
\end{equation}
where we have used \eqref{eq:psi} to introduce the constant $\phi = 1-\psi$ and rearranged the resulting terms to show the Einstein tensor $G_{\mu\nu}$ explicitly. 
Now, performing some integration by parts we get
\begin{equation} \label{var5}
\begin{split}
\delta S_\xi & = -\frac{1}{\kappa} \int d^4x\sqrt{-g}\, \xi^\mu\left( \nabla^\nu G_{\mu\nu} - \frac{1}{2}\nabla_\mu(\phi R) \right. \\ &  \left.- \frac{1}{2}\nabla_\mu(a^\rho\nabla_\rho R) - \kappa \nabla^\nu T_{\mu\nu}\right) +\, \textrm{ST}.  
\end{split} 
\end{equation}
Finally, taking cognizance of condition \eqref{eq:diffcond}, using the Bianchi identity  $\nabla^\nu G_{\mu\nu}=0 $, for an arbitrary $\xi^\mu$, and setting $\delta S_\xi $ to zero we get
\begin{equation} \label{eq:noncons}
\nabla_\mu(\phi R) + 2\kappa \nabla_\nu {T_\mu}^\nu= \phi \nabla_\mu R + 2\kappa \nabla_\nu {T_\mu}^\nu =0,
\end{equation}
which is exactly the result in the main text and indicates the non-conservation of the matter energy-momentum tensor.

\end{document}